%% file: arxiv.tex
\documentclass[10pt, letterpaper]{article}

\usepackage{cite}
\usepackage{amsmath,amssymb,amsfonts}
\usepackage{algorithmic}
\usepackage{algorithm}
\usepackage{graphicx}
\usepackage{textcomp}
\usepackage{booktabs}
\usepackage{multirow}
\usepackage{siunitx}
\usepackage{xcolor}
\usepackage{placeins} 
\usepackage{url}
\usepackage{array}
\usepackage{amsmath}
\usepackage{hyperref}
\usepackage{csquotes}
\usepackage[caption=false,font=normalsize,labelfont=sf,textfont=sf]{subfig}

\usepackage[normalem]{ulem} 
\usepackage{comment}

\def\BibTeX{{\rm B\kern-.05em{\sc i\kern-.025em b}\kern-.08em
    T\kern-.1667em\lower.7ex\hbox{E}\kern-.125emX}}

\makeatletter
\def\cpartline#1{\@cpartline#1\@nil}
\def\@cpartline#1-#2\@nil{%
  \omit
  \@multicnt#1%
  \advance\@multispan\m@ne
  \ifnum\@multicnt=\@ne\@firstofone{&\omit}\fi
  \@multicnt#2%
  \advance\@multicnt-#1%
  \advance\@multispan\@ne
  \hfill
        \leaders\hrule\@height\arrayrulewidth\hfill
        \leaders\hrule\@height\arrayrulewidth\hfill
        \leaders\hrule\@height\arrayrulewidth\hfill
  \hfill
  \cr
  \noalign{\vskip-\arrayrulewidth}}
\makeatother
    
\newcommand{\GDate}{2020/01/01}
\newcommand{\ASesCount}{9695~}
\newcommand{\IXPsCount}{686~}

\newcommand{\LinksCount}{52422~}

 \newcommand{\pc}{{\tt port\_capacity}\ }
 \newcommand{\ps}{{\tt port\_size}\ }
 \newcommand{\ir}{{\tt info\_ratio}\ }

\newcommand{\period}{2019-01-01 and 2021-03-01}
\newcommand{\periodA}{2020-01-01 and 2020-06-01}

\newcommand{\graph}{{\it pDB c-graph}}
\newtheorem{definition}{Definition}[section]

\graphicspath{{../fig/}}

\begin{document}
%

\title{
Global Internet public peering capacity of interconnection: a complex network analysis}

%

\author{
Justin~Loye \\
IRIT, LPT, Université de Toulouse \\
CNRS, Toulouse INP, UT3\\
Toulouse, France\\
justin.loye@irit.fr
\and 
Sandrine~Mouysset\\
IRIT, Université de Toulouse \\
CNRS, Toulouse INP, UT3\\
Toulouse, France\\
sandrine.mouysset@irit.fr
\and 
Marc~Bruy\`ere\\
IIJ, University of Tokyo,\\
CNRS IRL 3527\\
Tokyo, Japan\\
mbruyere@nc.u-tokyo.ac.jp
\and
Katia~Jaffr\`es-Runser\\
IRIT, Université de Toulouse \\
CNRS, Toulouse INP, UT3\\
Toulouse, France\\
katia.jaffres-runser@irit.fr
}


\maketitle

\begin{abstract}
A massive and growing part of Autonomous System (AS)-level traffic exchanges takes place at Internet Exchange Points (IXPs).
This paper leverages PeeringDB, a  database providing a partial but reasonable view of the global interconnection of ASes at IXPs, to model a complex graph enabling the characterization of the key Internet peering players and their interactions over time.
We model a PeeringDB snapshot as a weighted directed bipartite graph, called the \graph, that captures the port size ASes possess at IXPs using available metadata. This novel model of the Internet is shown to picture relevant features of a complex network that groups ASes and IXPs in geographical areas of influence. From this model, we extract central players of public peering such as hypergiant AS content providers and major regional traffic receivers. Most importantly, this graph model opens the way to apply spectral analysis using reduced Google matrix in order to retrieve the intensity of possible interactions between ASes on the basis of pure connectivity information. As an illustration, we retrieve the timely evolution of the peering network to show how the central content and cloud providers have increased their reach to eyeball networks during Covid-19 pandemic.   
\end{abstract}




\section{Introduction}
Internet is structured around Autonomous Systems (ASes), administered by Internet Service Providers (ISPs), content providers, universities, etc. They are located all over the world and connect at different Points of Presence (PoPs), either using a pairwise private peering connection or doing public peering in multi-party shared switch fabric known as Internet eXchange Point (IXP). In the early 2000's, most ASes connected through private links, with a customer/provider relationship inducing a hierarchical structure of the Internet. The last decade has seen a surge in the deployment and the growth of IXPs, leading to a flattened Internet topology \cite{GREGORI201168}. This momentum was driven by the more diverse and cheaper interconnection services IXPs offer around the world to reach regional delivery networks. Content providers find in IXPs a chance to reach in a relatively reduced deployment time, sometimes using a tier remote peering service, the regional networks connected to the end users \cite{bottger2018looking}.     
Fundamental works \cite{ager2012anatomy}\cite{Chatzis2013} showcased that IXPs were growing in importance by enabling more and more Internet traffic since the early 2010s. Recent studies showed that IXPs enabled to cope with the network stress caused by end-user increase demand during the Covid-19 outbreak \cite{NetflixOCA}\cite{feldmann2021year}.

Thus, knowledge of the public peering ecosystem and its underlying graph structure is a key element of understanding the global Internet topology. However to fully grasp the AS interconnections, one would need to obtain a traffic matrix for each IXP. Such traffic matrices are not publicly disclosed, and only a few have been described in the scientific literature in recent years\cite{cardona2012history}\cite{cardona2012ixp}\cite{ager2012anatomy}. To the best of our knowledge, the most comprehensive source of AS-level topology data is Caida's AS relationships\cite{giotsas2013inferring}. This dataset labels AS relationships as \enquote{customer-to-provider} and \enquote{peer-to-peer}. However peer-to-peer inference is notoriously hard and Caida's work, while invaluable to the research community, is known to miss most of those links. On the other hand, the authoritative source of information regarding public peering ecosystem is PeeringDB, a community database where ASes register their own information and IXP membership in order to find new peers. This IXP membership is best described by an unweighted and undirected bipartite AS-IXP graph \cite{nomikos2017re}, but such a graph does contain direct links between ASes and information about traffic exchanges.

To circumvent these problems, in this paper we define a weighted oriented bipartite graph model of the Internet based on PeeringDB. We leverage the  router \ps  that ASes have at IXPs to peer with other IXP participants. Actual traffic between ASes is not available, but we have an AS {\tt \ir} attribute that indicates the general tendency to receive or send traffic. We show that this bipartite graph has shared characteristics across complex networks and is expressive enough to encode the main features of the global peering ecosystem. With a novel spectral graph theory approach based on stochastic complementation from Markov chain theory\cite{meyer1989stochastic}, we leverage the graph to approximate AS-AS traffic exchanges at a global scale.
This new PeeringDB model allows us to make the following contributions:
\begin{itemize}
    \item identify the main sources of Internet traffic known as \enquote{hypergiants},
    \item extract the regional presence of ASes from the structural properties,
    \item identify the main traffic destinations as regional ISPs close to the end-user, known as \enquote{eyeball networks},
    \item study the diffusion patterns of hypergiants to regional ISPs with spectral graph theory, 
    \item illustrate the model on a longitudinal study of the Covid-19 demand for content.
\end{itemize}

The rest of the paper is organized as follows. We start by presenting PeeringDB in \autoref{sec:peeringdb}. We then introduce our graph model in \autoref{sec:model} and perform an analysis of its main features in \autoref{sec:analysis}. In \autoref{sec:regomax}, we introduce a graph theory tool known as stochastic complementation \cite{meyer1989stochastic} to study the hypergiants diffusion to eyeball networks. Lastly, we show in \autoref{sec:covid} how hypergiants reacted to the demand increase from end-users during Covid-19 outbreak.

All processed data, source codes and network visualisation files will be made available \cite{loye2022peeringdb} for final publication. 

\section{PeeringDB}
\label{sec:peeringdb}
\begin{figure}
    \centering
    \includegraphics[width=4in]{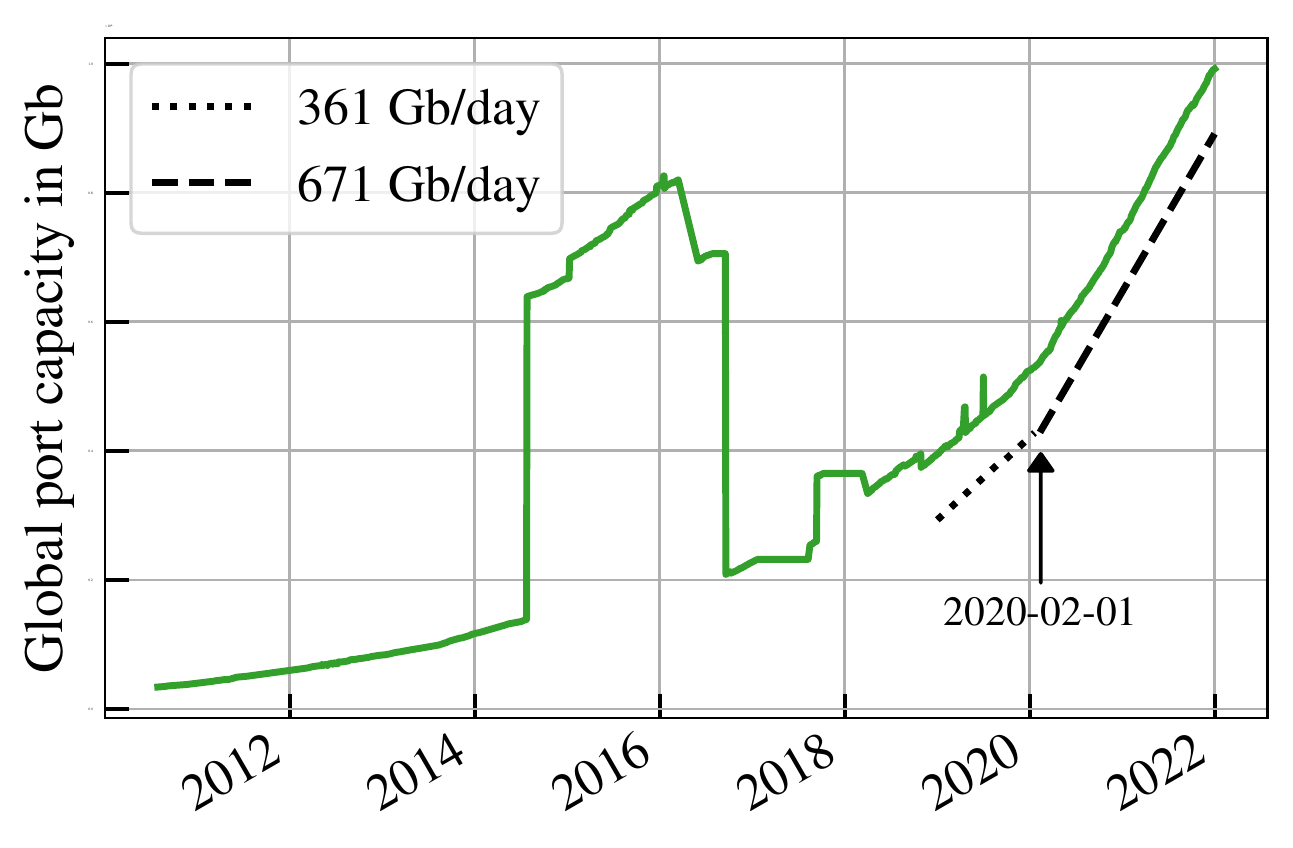}
    \caption{Daily evolution of the total port capacity of ASes of PeeringDB between 2010-07-29 and 2021-12-31. {\it An inflection point in 2020-02 is underlined by two linear regressions with respective slope of 361 and 671 Gbit/day.}}
    \label{fig:portcapa_tot}
\end{figure}
PeeringDB\footnote{https://www.peeringdb.com/} is a non-profit and user-maintained database that facilitates the global interconnection of networks at IXPs, data centers, and other interconnection facilities. Network operators register information about their organisation, networks and point of presence, creating a picture of the peering ecosystem. From this picture, network operators are able to identify new potential peers and how to choose the best PoP to expand their network. As of 2022, 12689 registered ASes reports presence at IXPs. In total, there are 872 IXPs reporting members, of which 771 have at least 3 participants and can thus be used for multilateral traffic exchange. PeeringDB can be accessed publicly with an API, and daily dumps are made available to the community by Caida from 2010 to the present day \cite{lodhi2014}.

PeeringDB is recognized as an authoritative source of information. Most of the data is now uploaded automatically using IXP management software. For instance, small to medium size IXPs generally deploy {\it IXP Manager} \footnote{\url{https://www.ixpmanager.org/}} which is currently used by 184 IXPs. Larger IXPs have as well automated PeeringDB declarations in their in-house management software through the PeeringDB API. As such, network operators can use PeeringDB as part of their automation process. It is in an organisation's best interest to have an  accurate and updated record.
PeeringDB has been validated by BGP data in \cite{lodhi2014}, where authors show that PeeringDB membership is generally up-to-date and reasonably representative of the Internet's transit, content and access providers in terms of business types and geography of participants.
Following a more recent study from \cite{bottger2018looking}, we argue that PeeringDB can be used as a ground-truth source of data of the publicly declared part of the Internet.

In this paper, we will make use of the port sizes ASes possess at IXPs, reported in PeeringDB, indicating an upper bound to the actual real-time transit taking place between ASes. The sum of all port sizes called port capacity, computed for each AS, has been shown to be a discriminating feature to identify hypergiants ASes\cite{bottger2018looking}. The present work focuses on ASes and IXPs data contained in PeeringDB snapshots collected between 2019-01-01 and 2021-03-01. Before this period, snapshots are more prone to discrepancies that are reflected as outliers in port capacity. The most striking example, also reported in another study \cite{jageranalyzing}, is the 2014-2016 bump in global port capacity presented in \autoref{fig:portcapa_tot}. This bump was caused by a single Australian AS, connected at 5 IXPs, that reported port sizes one hundred times bigger than Akamai, a global content delivery network. From 2019-01-01, outliers are fewer and do not persist over time, indicating efforts in data curation by PeeringDB. In the following, we made sure that the snapshots we are building our conclusions on are not impacted by outliers.

\section{A new PeeringDB modelling}
\label{sec:model}
ASes membership to IXPs reported in PeeringDB can be naturally represented as a weighted bipartite graph. This allows to take advantage of the powerful tools from graph theory, as we will see in \autoref{sec:regomax}.
Motivated by this, we introduce a novel graph model, called  the \graph, on the basis of the model defined by Nomikos et al. in \cite{nomikos2017re}, but augmented with both weighted and directional edges. 

\subsection{Graph model}
The \graph~ is constructed as a bipartite graph, where the set of vertices ${\mathcal V}$ is split between two sets $\mathcal A$ and $\mathcal X$ representing the ASes and IXPs of a PeeringDB snapshot. Edges exist between an AS $as \in \mathcal A$ and an IXP $ix \in \mathcal X$ if $as$ is a member of $ix$.  Edges $\mathcal E$ are weighted with the router \ps metadata, which is associated to inbound and outbound directions following ASes \ir metadata.
We associate to each undirected edge $e_{i\leftrightarrow j} \in {\mathcal E}$ a real number $ps(e_{i\leftrightarrow j})$ corresponding to the \ps. If an AS has multiple routers at an IXP, we agglomerate their port size to weight a single link. Since the graph is bipartite, $ps(e_{i\leftrightarrow j}) = 0,  \forall (i,j) \in {\mathcal A} \times {\mathcal A} ~\lor~ \forall (i,j) \in {\mathcal X} \times {\mathcal X}$.
The \graph~ is thus defined as:
\begin{definition}[\graph]
\begin{equation*}
    {\mathcal G} = ({\mathcal E}, {\mathcal V}), {\rm ~~with~~} {\mathcal V} = {\mathcal A} \cup {\mathcal X} 
\end{equation*}
\end{definition}

The edges direction is derived from both \ps and \ir metadata as follows. In PeeringDB, each AS is associated with the \ir label representing the traffic imbalance in its relationship with other IXPs' members. Thus, traffic at an AS can be reported as Heavy Outbound (HO), Mostly Outbound (MO), Balanced (B), Mostly Inbound (MI), Heavy Inbound (HI) or Not Disclosed. Around 83\% of ASes have a valid label as presented in Table~\ref{tab:info_ratio_tab}, these ASes representing around 94\% of the total \pc. 
\begin{table}[!t]
    \caption{Proportions of ASes in size and \pc for \ir categories.}
    \centering
    \input{info_ratio_tab}
    \vspace{3pt}
    \label{tab:info_ratio_tab}
\end{table}

We leverage this qualitative information to derive a specific port size value for {\it directed} inbound and outbound edges. Since links are full-duplex, we assume maximum capacity is used by the AS in the direction declared in \ir label. This hypothesis is reasonable since we model the capacity provisioned by ASes at IXPs and not the real traffic volumes. Therefore, we define a weighted and directed adjacency matrix $W$, with elements $W_{ij}$ indicating a link from $j$ to $i$, the following way: 

\begin{definition}[Weighted and directed adjacency matrix $W$]
\noindent Let $i \in \mathcal{A}$ and $j \in \mathcal{X}$.
If $i$ is registered inbound
\begin{align}
W_{ij} &= ps(e_{i\leftrightarrow j}),\hspace{1.5cm} W_{ji} = \left(1-\beta\right) \cdot ps(e_{i\leftrightarrow j}),
\label{eq:edge_orientationIn}
\intertext{and if $i$ is registered outbound}
W_{ij} &= \left(1-\beta \right) \cdot ps(e_{i\leftrightarrow j}),\hspace{0.72cm} W_{ji} = ps(e_{i\leftrightarrow j}),
\label{eq:edge_orientationOut}
\end{align}
with a $\beta \leq 1$ coefficient set for either {\tt Balanced} ($\beta_B$), {\tt Not Disclosed} ($\beta_{ND}$), {\tt Heavy} ($\beta_H$) or {\tt Mostly} ($\beta_{M}$) traffic imbalance classes of ASes. 
\end{definition}

A default parameter setting is considered in this paper:
\begin{itemize}
    \item[-] {\tt Balanced} ; {\tt Not Disclosed} : $\beta_B = 0$
    \item[-] {\tt Mostly} : $\beta_M = 0.75$
    \item[-] {\tt Heavy} : $\beta_H = 0.95$
\end{itemize}
An edge $e_{i\leftrightarrow j}$ is thus decomposed in two directed edges $e_{i\rightarrow j}$ and $e_{i\leftarrow j}$ of different weight. 
If $\beta_B = 0$, Eq.~\eqref{eq:edge_orientationIn} or Eq.~\eqref{eq:edge_orientationOut} are equivalent, and each directed edge is being assigned the full \ps in weight. 
With the default setting, a heavy (resp. mostly) outbound AS is connected to its neighbors with outgoing edges of weight equal to the full port size, and with incoming edges of weight equal to 5\% (resp. 25\%) of the port size.

Our model captures capacities of provisioned links in the Internet, which is of course different from the traffic exchanges occurring in reality. It corresponds to an upper limit of real traffic. In the following, we use this loose upper-bounded definition of traffic to describe the provisioned bandwidth of Internet players. Thus, every time we use the term `traffic' we refer to the oriented link capacities of $W$.  
\subsection{Parameters and network stability}
Parameters $\beta_H$ and $\beta_M$ control the capacity imbalance between outbound and inbound edges. The \graph~ would benefit from a non-uniform personalized $\beta$ to capture real world traffic imbalance but such data is rarely disclosed. We keep thus a uniform value of $\beta_H$ and $\beta_M$, and question here their influence on the network topology. 
To do so, we generate 400 networks for 20 equally-spaced values of $\beta_H\in[0.9, \dotso,1]$ and of $\beta_M\in[0.6, \dotso, 0.8]$. To measure the impact of these parameters, we select for each \ir category the 4 ASes that rank best in {\tt port\_capacity}, and for these ASes, we record their centrality in the network with PageRank (PR) \cite{BrinPage98} and reverse PageRank (rPR)\cite{Bar-Yossef08}. PageRank is a measure of nodes importance in terms of the weighted incoming links, in our case the ability of a node to capture traffic from the rest of the network. The ability of a node to disseminate traffic is also of interest and is obtained by computing PageRank on the same graph but with all links inverted. This centrality is known as reverse PageRank.

 We observe that $\beta_H=1$ causes a disruption since HO ASes like Facebook or Netflix have no incoming traffic, resulting in ranking them last in PageRank. This situation being not realistic, we exclude the corresponding networks from the results of Table \ref{tab:stability_tab}. This table lists PR and rPR values of the 16 ASes when the network is constructed with default parameters $(\beta_H=0.95, \beta_M=0.75)$, and most importantly, the maximum variation of these values over the $\sim400$ graphs. The lower these variations, the less sensitive PR and rPR centrality metrics are to the choice of $\beta_H$ and $\beta_M$.

 \begin{table}[!t]
    \caption{AS rank stability to model parameters. {\it PageRank (PR) and reverse PageRank (rPR) values correspond to $(\beta_H=0.95, \beta_M=0.75)$. $\Delta PR$ and $\Delta rPR$ correspond to the variation of PR and rPR observed for the $19 \times 20$ graphs created with $(\beta_H, \beta_M) \in [0.9, \dotso,.1[ \times [0.6, \dotso,0.8]$. IR column represents the \ir label following the notation of Table~\ref{tab:info_ratio_tab}. We use the 2020-01-01 snapshot with \graph~ of size $N=10381$.}}
    \centering

\input{stability_tab.tex}
    \vspace{3pt}
\label{tab:stability_tab}
\end{table}{}

First, balanced ASes are very stable in rPR and PR as expected.   
Our results show that outbound ASes are stable with respect to rPR and inbound ASes are stable with respect to PR. For these ASes, the default parameters create a representative network. Conversely, outbound ASes have a high $\Delta$PR and inbound ASes a high $\Delta$rPR. This last observation shows that we can only analyze outbound ASes with rPR metric and inbound ASes with PR metric to keep results calculated with the default parameter set representative. Thus, in the rest of the paper, \graph~ is constructed with default parameters, and all investigations related to outbound (resp. inbound) ASes are made with rPR (resp. PR) metric.

\subsection{Traffic balance at IXPs}

In order to verify the consistency of PeeringDB data, we study for each IXP the traffic imbalance induced by the \ps of its participants. By definition, traffic at IXPs is balanced between inbound and outbound flows. Since our graph model builds on the \ps and \ir reported by ASes, exhibiting balanced traffic at IXPs asserts the consistency of PeeringDB data and shows that our graph holds as a reasonable model of Internet peering.

\begin{figure}
    \centering
    \includegraphics[width=0.7\textwidth]{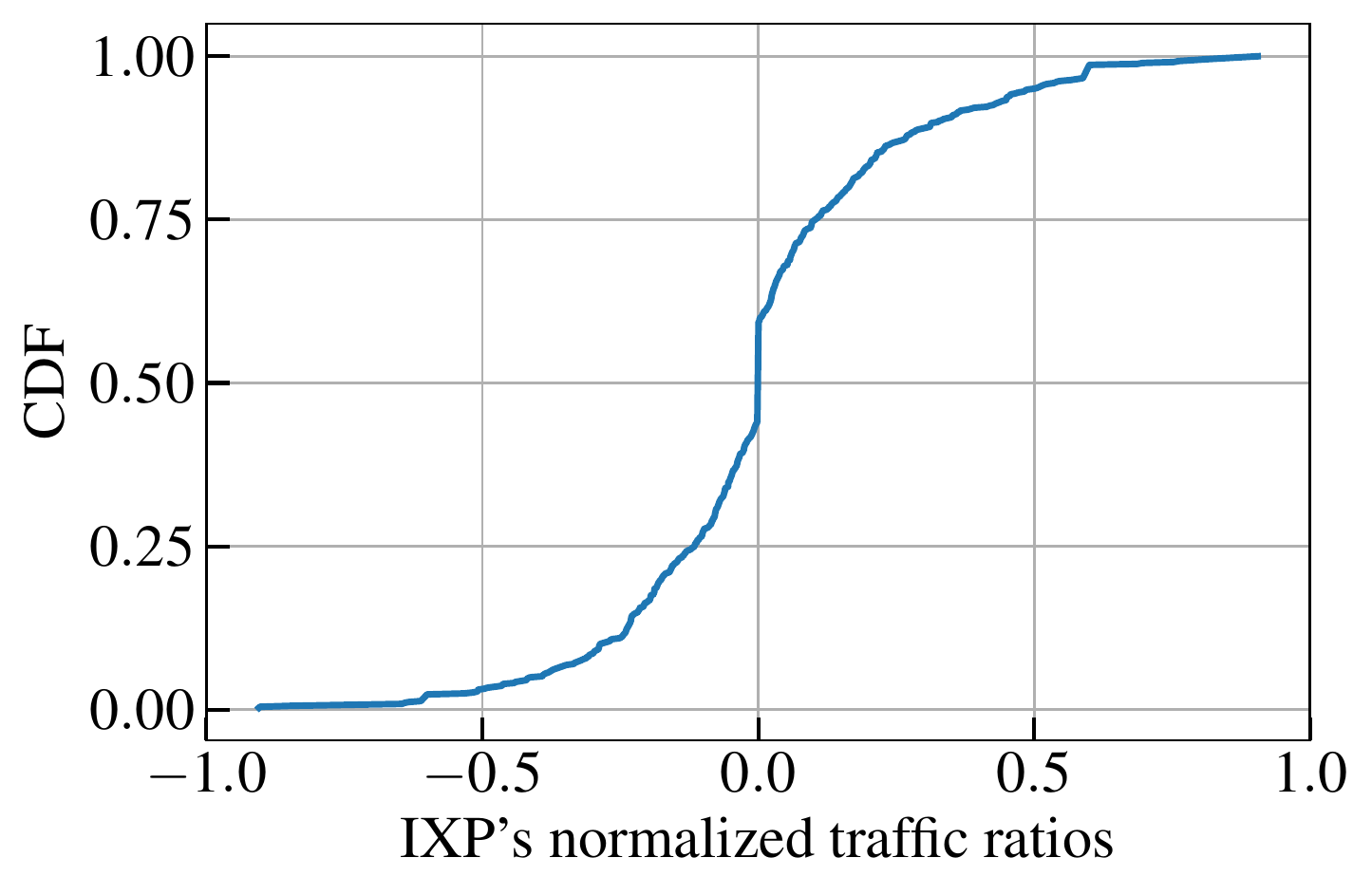}
    \caption{IXPs normalized traffic ratios or balance $B$.}
    \label{fig:ixps_ratio_cdf}
\end{figure}

The \pc metric is directly related to our \graph~ definition as follows. Let $w^i_\mathrm{in}=\sum_{j}W_{ij}$ be the weighted incoming degree and $w^i_\mathrm{out}=\sum_{j}W_{ji}$ the weighted outgoing degree.
For an inbound AS $i$, it is clear from \eqref{eq:edge_orientationIn} that  
\begin{align}
w^i_\mathrm{in}&={\tt port\_capacity}(i),   w^i_\mathrm{out}=(1-\beta).{\tt port\_capacity}(i).\nonumber\\
\intertext{In the same way, for $i$ outbound, }
w^i_\mathrm{in}&=(1-\beta).{\tt port\_capacity}(i),  w^i_\mathrm{out}={\tt port\_capacity}(i). \nonumber
\end{align}
But these relations do not hold for IXPs since \pc is induced by the membership of ASes. Therefore, we can assert the correctness of AS reported \ir and by extension the consistency of our model in studying the normalized traffic balance at IXPs given by  $B(i) = (w^i_{\mathrm{out}}-w^i_{\mathrm{in}})/(w^i_{\mathrm{out}}+w^i_{\mathrm{in}})$. This balance ranges from -1 to 1, with 0 representing a balanced IXP. 
We show in \autoref{fig:ixps_ratio_cdf} that IXPs are overall balanced with a mean $\langle B\rangle=0.00\pm0.26$ and respective quartiles (Q1, Q2, Q3) at (-0.12, 0.00, 0.10). These results show that, at the local scale of IXPs, PeeringDB data gives a consistent view of AS membership, port size and \ir metadata. 

\section{Leveraging complex Internet network analysis}
\label{sec:analysis}
The goal of this section is to gain insight into the peering ecosystem solely from the \graph~ structure. We first present a general overview of the graph that underlines the complex network nature of PeeringDB, reporting on topology metrics such as degree and weight distributions, and the relationship between \pc and node degree of ASes. Motivated by the fact that most of today's Internet traffic flows from hypergiants to regional eyeball networks, we show in a second part how to recover these key peering actors in an unsupervised manner.
\label{sec:pdbcgraph}

\subsection{Overall graph description}
\label{subsec:graph_description}
As of \GDate, it consists of \ASesCount ASes that are linked to \IXPsCount IXPs by \LinksCount links. Over 99\% of all nodes belong to the largest connected component of the graph
\subsubsection{Impact of weighting the edges}
First of all we illustrate the benefits of weighting the bipartite graph with directional \ps by comparing the PR and rPR centralities of our model to the ones obtained with the non-weighted non-oriented IXP graph of Nomikos et al.\cite{nomikos2017re}. For the \graph, differentiating the \ps on the oriented edges offers a clear distinction between the role of nodes. A node with high PR (rPR) is likely to receive (send) a large amount of traffic. Results for PR and rPR are shown in Table~\ref{tab:PRCR_comp_tab}: top-15 PR nodes are essentially IXPs that represent the authorities in our \graph, while top-15 rPR identify hypergiants that diffuse their content in the network. Such a clear distinction is not present in a network model where edges are neither weighted nor oriented. For instance, an undirected graph can't distinguish the strong diffusion of content providers like Facebook, Amazon, Netflix or Akamai from the central role of large IXPs. 

\begin{table}[t]
    \caption{Comparison of the top 15 PageRank and reverse PageRank of IXP/AS network using our \graph, and the model of \cite{nomikos2017re}. {\it Graphs are constructed for 2020-01-01 ; ASes are highlighted in bold font.}}
        \centering

\input{PRCR_comp_tab.tex}
    \vspace{3pt}
\label{tab:PRCR_comp_tab}
\end{table}

\begin{figure}[!t]
    \centering
    \includegraphics[width=3.3in]{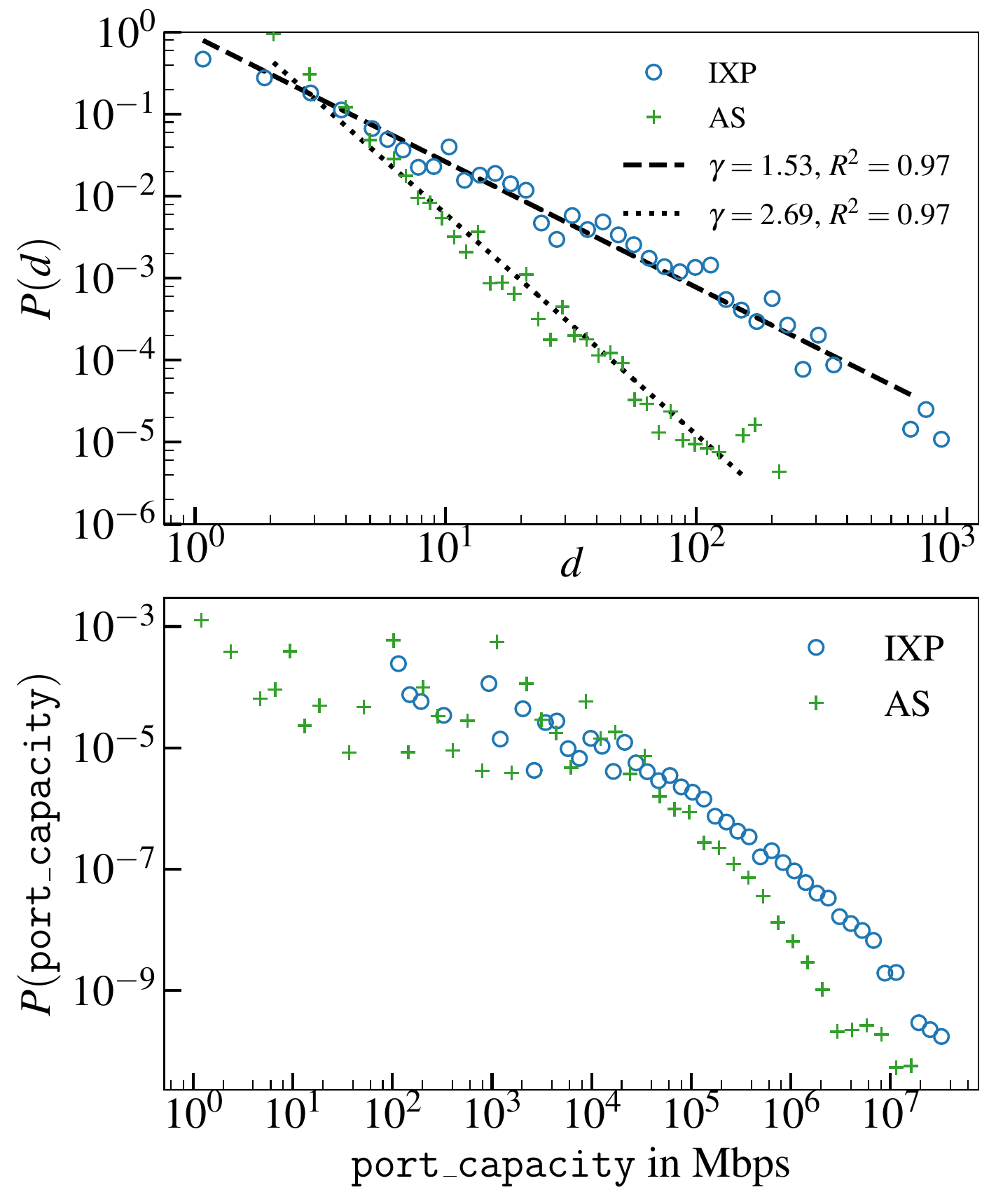}
    \caption{\graph~ degree and weight distributions. {\it Degree distributions follow a power-law $f(x)=Ax^{-\gamma}$}.}
    \label{fig:power_law}
\end{figure}

\begin{figure}[!t]
    \centering
    \includegraphics[width=3.3in]{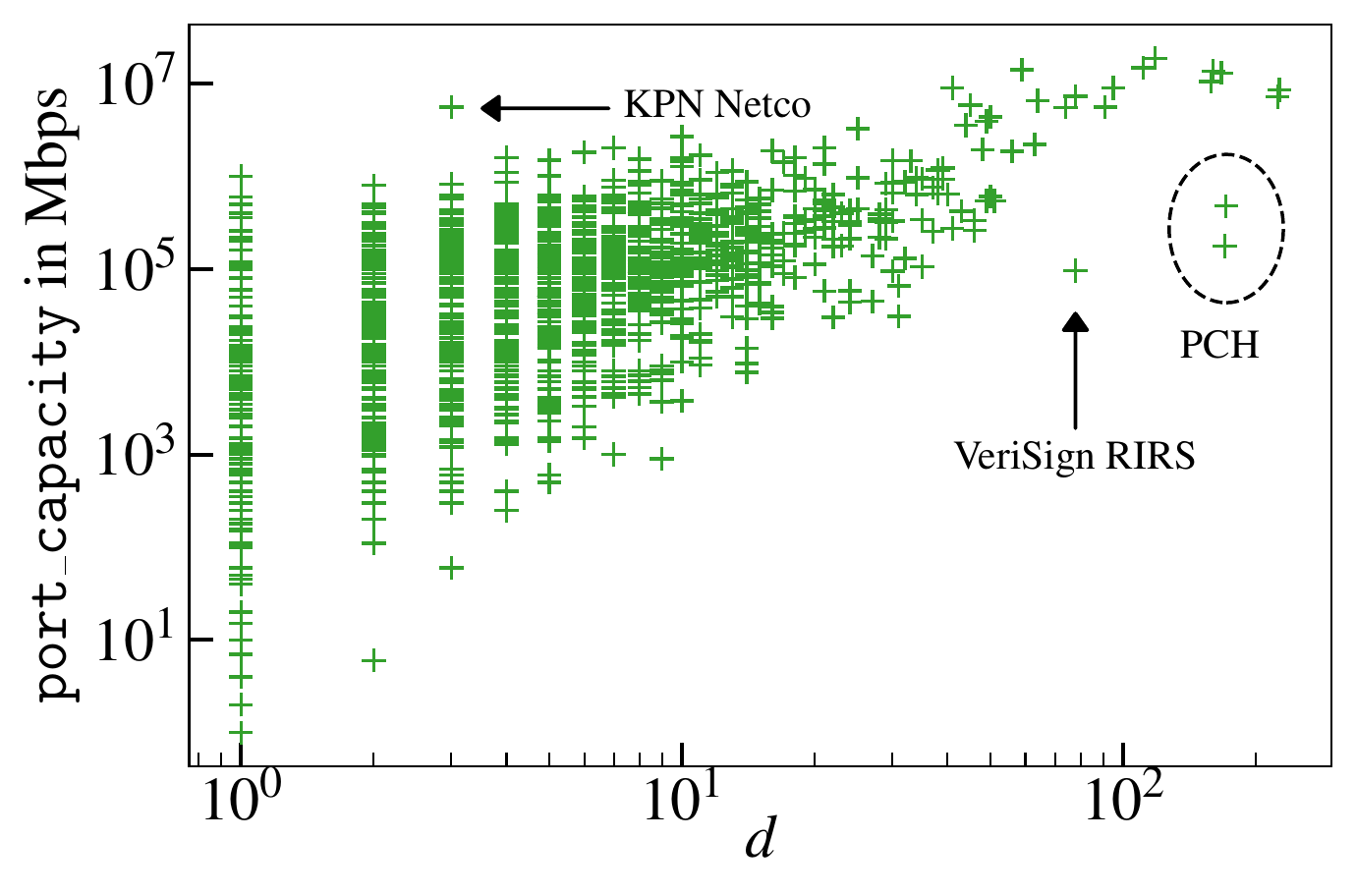}
    \caption{\pc versus node degree $d$ for ASes.}
    \label{fig:pc_vs_d}
\end{figure}

\begin{figure*}[h]
    \centering
    \hrule
    \includegraphics[width = 6.1in]{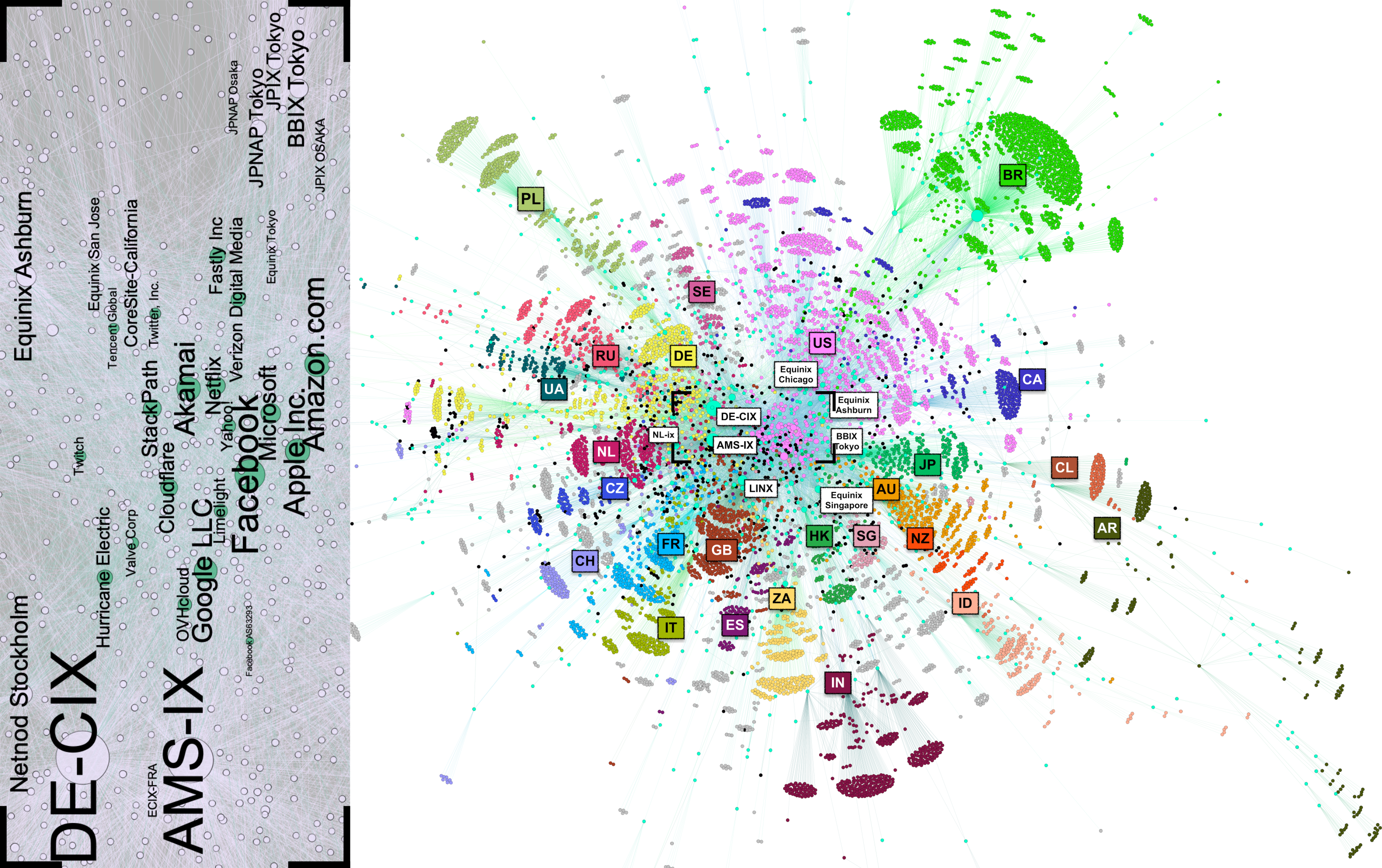}
    \hrule
\caption{\graph: Visualization of ASes classified by countries plot using OpenOrd and Yifan Hu layouts of Gephi \cite{gephi}). {\it 
Node size is proportional to \pc and cluster colors are indicated with country label colors for a selection of the largest 25 clusters. IXP node color is light blue and ASes identified as {\tt tied} are black. On the left, an inset figure zooms on the position of the 20 hypergiant content providers, together with the largest IXPs, located at the center of the graph.
Graph can be navigated interactively at \cite{viz_network_topology}.}}
    \label{fig:clusters}
\end{figure*}

\subsubsection{Degree and \pc distributions}
Probability distributions of undirected nodes degree $d$ and port capacity for the \graph\ 
are shown in \autoref{fig:power_law}.
From our model definition, this degree is equal to the degree of incoming links and to the degree of outgoing links.

The probability distribution of node degree is generally well approximated by a power-law, a feature commonly found in other real networks \cite{barabasi1999emergence}. Power-law exponent of ASes degree distribution $\gamma = 2.69$ falls in the range $[2,3]$ usually reported for real networks \cite{clauset2009power} and shows that the repartition of ASes degree is heterogeneous, especially compared to the IXPs, whose degree distribution is fitted with $\gamma = 1.53$. \pc distribution follows a heavy-tailed distribution found in other weighted networks such as the world-wide airport traffic network \cite{barrat2004architecture}, where authors introduce a \enquote{node strength} metric analogous to our {\tt port\_capacity}. 

As opposed to the latter network of \cite{barrat2004architecture}, we do not find in \autoref{fig:pc_vs_d} a particularly linear relation between AS degree and {\tt port\_capacity}. We attribute this observation to the fact that port size may vary from several orders of magnitude (100M to 1T), while airplanes carry tens to hundreds of passengers. Interestingly, we find outliers that present different peering strategies. Regional Network Service Provider KPN-Netco ranks 15 in \pc with only $d=3$. On the other hand, ASes that support public Internet services (e.g. DNS) such as both Packet Clearing House (PCH) ASes and VeriSign Global Registry Services respectively rank according to the degree 3rd and 4th for PCH, and 13th for VeriSign, with a \pc of about two orders of magnitude lower than the hypergiants that exhibit similar values of degree.

\subsection{Retrieving key peering actors}
We now investigate the structural properties of the graph in order to retrieve the main peering actors. We first start by deriving a country classification of ASes and proposing a visualization of the graph. We identify hypergiants with the reverse PageRank network metric, and show with the visualisation their global reach. Then for countries of interest we extract the main traffic receivers with PageRank metric, and show that identified ASes relate to eyeball networks.

\subsubsection{\graph\ regional structure}
\label{subsection:clustering}
We aim at refining the geographical information reported by ASes in PeeringDB under the label \texttt{info\_scope}. This attribute can either be referenced with the name of a continent or as {\tt Global}, {\tt Regional}, {\tt Not Disclosed}. Refining allows to narrow down ASes location from continent-level to country-level, which is particularly needed since 45\% of ASes are reported as {\tt Regional} or {\tt Not Disclosed}. 
In order to attribute countries to ASes, we refer to their relationships with IXPs. Indeed, IXPs have accurate {\tt country} labels. Therefore, we assign to an AS the country of the majority of IXPs it belongs to. For 5.46\% of ASes there is no majority, and we give them the label {\tt Tied}.

 \begin{table}
    \caption{ASes country classification metrics. {\it Ground-truth data is obtained from \cite{ASorg}, and prediction is based on the country of the majority of IXPs an AS belongs to.}}
    \centering
    \input{geographical_classif_tab}
\label{tab:geographical_classif}
\end{table}{}

The geographical classification has been assessed by Caida AS organizations dataset \cite{ASorg}. This dataset maps ASes to organizations, and retrieve organizations' country with regional Internet registries WHOIS entry or by inference. By considering this dataset as ground-truth, classification metrics such as precision, recall and F1-score are computed and 
presented in \autoref{tab:geographical_classif}. Recall is defined as the ratio of the true positives to the union of true positives
and false negatives and precision is the ratio of the true positives to the union of true positives
and false positives.  High recall minimizes false negatives while high precision minimizes false positives. Finally, F1 score is an harmonic mean of precision and recall. High F1 score occurs if both recall and precision are
high.
We confirm that overall, country classification based on ASes' proximity to IXPs in PeeringDB is in good accordance with Caida.

We question if this geographical classification is consistent with the structure of the \graph. To answer this question, we propose a visualization of the \graph\ of 2020-01-01 in Figure~\ref{fig:clusters}. The graph layout, computed with the automatic Yifan Hu force-direct algorithm of Gephi \cite{gephi}, is available to be interactively navigated in \cite{viz_network_topology}. The geographical distribution of ASes is in good accordance with the \graph\ structure.
Links and their weights arrange the nodes such as to structure the graph around areas of influence of IXPs, which tend to be correlated with their geographical location. The ASes identified as {\tt Tied} are spread evenly across the network and are often peering at a few IXPs of different countries.
Interestingly, the graph topology not only groups ASes per countries, but keeps the geographical proximity of countries. For instance, the upper left part of the graph groups Northern and Eastern European countries, while the lower left groups Central European ones. The lower right part groups Asian and Oceanian countries, while the upper right groups the USA, Canada and Southern-American countries. This relation between regional and structural proximity is also found with Louvain method, an unsupervised clustering technique presented in Appendix \autoref{app:louvain}.


\subsubsection{IXPs and hypergiants global reach}
\begin{table}[!t]
    \caption{Hypergiant ASes: Top 20 ASes in reverse PR in \graph~of 2020-01-01. {\it ASes identified as hypergiants in \cite{bottger2018looking} are listed in bold font.}}
    \centering
    \input{hypergiants_tab.tex}
    \vspace{3pt}
\label{tab:hypergiants_tab}
\end{table}{}
We extract the best 20 highly diffusive ASes with reverse PageRank metrics. These ASes, listed in \autoref{tab:hypergiants_tab}, encompass the 15 hypergiant ASes of \cite{bottger2018looking}, most of them being content providers. In the rest of this paper, we identify this set as the set of hypergiants. 
In \autoref{fig:clusters}, highlighted with an inset view, we find hypergiants and important IXPs that play a central role since they have a very large number of links. 
For instance, content providers such as Cloudflare, Akamai, Microsoft, Facebook, Amazon or Netflix are respectively connected to 226, 160, 158, 118, 111 or 95 IXPs. Hurricane Electric, PCH or VeriSign are connected to 224, 170 or 78 IXPs. 
Central IXPs like DE-CIX Frankfurt and AMS-IX interconnect the hypergiants to European ASes, while Japanese connect them to Asia and Oceania. US-based Equinix IXPs link the US ASes to the rest of the network.  
The set of ASes connected to a single IXP is represented with an umbrella shaped form. These ASes are mostly of {\tt Cable/DLS/ISP} type. The closer to the center an AS is, the higher its degree gets as it is connected to multiple IXPs. We observe as well that higher degree ASes are tagged with {\tt NSP} (Network Service Provider) or {\tt Content} type in majority. This last observation, explored further in the next sections, reveals a geographic proximity that is leveraged by hypergiants to reach global and regional NSPs and ISPs through peering exchanges points. 
\subsubsection{Eyeball networks}
\label{sec:eyeball}
\begin{table*}[!t]

    \caption{ASes end-users market share by country. {\it The AS selection differs, from the first to the last line: all APNIC ASes, ASes present in both PeeringDB and APNIC, the top4 ASes by EUMS present in both PeeringDB and APNIC, the top 4 PR ISP and Not Disclosed ASes, traffic receivers (defined in the body).}}
    \centering
    \input{coverage_horizontal}
    \vspace{3pt}
\label{tab:coverage}
\end{table*}{}

We saw that hypergiants are easily identified with \graph.
Now that we know from where most of the traffic originates, we question here if ASes closest to end-users, known as eyeball networks, can be retrieved from PeeringDB. Intuitively, we expect these networks to be characterized by a strong inbound behavior in their geographical area of influence. As such, we will identify them using PR metric and the geographical information of  \graph. 

For the validation, we consider the APNIC customers per AS datasets \cite{apnic}. APNIC, via ad-based measurements, estimates for each eyeball ASes their end-user market share (EUMS) by country. We check first if APNIC eyeballs are covering, for 15 countries of interest, 100\% of the population. We see in \autoref{tab:coverage}, APNIC ASes entry, that the agglomerated EUMS is generally close to 100\%. However, if we select ASes that are both in APNIC and PeeringDB (cf. line 2 of table \autoref{tab:coverage}), this percentage drops more or less dramatically, depending on countries. The most underrepresented countries are India, Poland and the USA which underlines that eyeball ASes in these countries can't or don't leverage IXPs. Another limitation is that some eyeball ASes do not benefits from public peering, and therefore are not present in \graph. For example ISP Comcast, with 15\% market share in the US, does not report membership in public peering exchanges as a consequence of a paid peering policy \cite{DrPeeringComCast}.

We expect eyeball ASes to have an inbound nature and a regional presence visible in the structure of \graph. To identify them, we select for each country a subset of ASes with reported business type \texttt{DSL/Cable/ISP} or \texttt{Not Disclosed}. In this subset, we retrieve the top 4 traffic receivers, i.e.\ ASes that rank in the top 4 according to PageRank. We exclude from this set the ASes already present in the hypergiants set. We check if these ASes are present in APNIC and retrieve their EUMS (\autoref{tab:coverage} entry Top4 PR ISP/ND). We compare this number to the best accessible EUMS, i.e.\ the agglomerated EUMS of the top4 ASes present in both APNIC and PeeringDB (\autoref{tab:coverage} entry PeeringDB and APNIC top4). For 6 countries highlighted in bold, our procedure recovers more than half of the best accessible EUMS, showing that \graph\ can recover some regional eyeball networks. We can improve our results by considering NSPs as well as ISPs and then manually remove Not Disclosed or NSPs ASes that do not have end-user customers. We call ASes identified with this procedure \enquote{traffic receivers}. The aim is not to optimize the EUMS but rather to remove obvious wrongly identified ASes with as little alteration as possible. We have removed Verisign, Telegram Messenger, and the 4 tier-2 ISPs: Core-Backbone, Open Peering, Orange Polska and Brightwave. We observe that Canada, Russia and China eyeballs are not well recovered, mainly because of our clustering procedure miss-assigning them to the US. For other countries, we identified networks that reach a significant part of the population.

The main regional traffic receivers identified by our procedure are shown in \autoref{tab:catchers_tab}. We see that PeeringDB derived country clustering is in good accordance with APNIC countries\footnote{For an AS with presence in multiple countries, we selected the country where the AS has the largest EUMS.}. A notable mislabeling is the attribution in the US of NSP KDDI, representing 20\% of Japan EUMS. Almost all ASes identified are present in APNIC. For most countries, we are able to retrieve at least one or two ASes with a high ranking in APNIC.
In the next parts, we will characterize how these eyeballs capture hypergiants' content.


\begin{table}[]
    \caption{Main regional traffic receivers. {\it ASes are retrieved with the procedure described in \autoref{sec:eyeball}.}}
    \centering
\input{catchers_apnic_tab}
    \vspace{3pt}
\label{tab:catchers_tab}
\end{table}{}

\section{Deriving AS-AS traffic exchanges to study hypergiants diffusive patterns}
\label{sec:regomax}
ASes engages peering sessions at IXPs in order to exchange traffic. To fully capture these interactions, one would have to obtain a traffic matrix for each IXP.
Such traffic matrices are not publicly disclosed, and only a few of them has been described in the scientific literature over the past years \cite{cardona2012history}\cite{cardona2012ixp}\cite{ager2012anatomy}.
Capturing AS to AS traffic exchanges occuring at IXPs is therefore not achievable at a global scale and within a short-time resolution of a day.

In this section, we show that the graph formulation of PeeringDB, coupled with stochastic complementation from Markov chain theory, can partially address this problem. In simple worlds, we rely on co-ocurrence of ASes at IXPs, weighted by the port sizes and ASes reported ratio between inbound and outbound traffic, to approximate ASes traffic exchanges at a global scale. Our method consists in two steps i) building a Google matrix that encodes interactions of \graph\ ii) computing a reduction of this matrix, called the reduced Google matrix, for only nodes of interests. The matrix reduction allows us to censor IXPs and retrieve only indirect AS-AS interactions. We will use it to study the diffusive patterns of hypergiants toward the main regional traffic receivers identified in last section. In the rest of the paper, we refer to the {\it reduced network of ASes} as the union of hypergiants and regional eyeballs listed in both  Table~\ref{tab:hypergiants_tab} and \autoref{tab:catchers_tab}. 

\subsection{Node censoring and graph reduction}
\subsubsection{Google matrix}
From the weighted oriented matrix of a network of size $N$, the Google matrix $G$ is a stochastic matrix defined by 
$G_{ij}=\alpha S_{ij}+(1-\alpha)/N$
with $S$ the Markov transition matrix whose elements follow: 
$$S_{ij}=
\begin{cases} 
W_{ij}/w^j_{\mathrm{out}} \text{ if}~ w^j_{\mathrm{out}}>0, \\
1/N \text{ otherwise.}
\end{cases}
$$
The damping parameter $0\leq\alpha<1$ insures the irreducibility of $G$, that is for any normalized vector $P$ the iterative product $P=GP$ converges to a stationary distribution called the PageRank vector. Standard value for $\alpha$ is 0.85 as given in the state-of-the-art work of Brin and Page \cite{BrinPage98}. $P(i)$ is the probability of finding a random surfer at node $i$ if we let him surf the graph for an infinitely long time following the Markov transition probabilities of $G$. Thus sorting $P$ by descending values ranks the nodes from the most influential nodes in terms of incoming links to the less influential node. This ranking is called the PageRank centrality PR. Reverse PageRank is calculated by simply transposing $W$, which resumes to inverting all the links in the graph.
\subsubsection{Reduced Google matrix}
The reduced Google matrix describes the properties and interactions of a given subset of nodes belonging to a much larger directed network \cite{frahm2016wikipedia}. It is derived using a dedicated algorithm which is based on the stochastic complementation \cite{meyer1989stochastic} of the Google matrix. 

For a subset of $N_r$ nodes of interest selected in ${\mathcal G}$, we can define a reduced network by reordering rows and columns of $G$ in the following block structure: 
$$\widetilde{G}=\begin{bmatrix}
G_{rr} & G_{rs}\\
G_{sr} & G_{ss}
\end{bmatrix}$$
where the index $r$ refers to the nodes of the reduced network and $s$ to the nodes in the complementary network.
By noting the Pagerank vector as
$[P_r, P_s]^T$
that satisfies $\widetilde{G}P=P$, we define the {\it reduced Google matrix}, noted $G_R$, as $G_RP_r=P_r$. From the block structure of $\widetilde{G}$, $G_R$  can also be defined as the Schur complement of the block $G_{ss}$ with:
\begin{equation}
G_R=G_{rr}+G_{rs}(\textbf{1}-G_{ss})^{-1}G_{sr}. \label{GR}
\end{equation}
The matrix $G_R$ represents an equivalent Markov chain whose transition probabilities capture the interactions between the subset of selected nodes by direct and all possible indirect (multi-hop) links present in the network of origin.

\subsection{Hypergiants diffusive patterns}
To study hypergiants diffusive patterns, we construct a reduced Google matrix encompassing them and the main regional traffic receivers. The benefits of the reduction are two-fold. First, by censoring the IXPs, we obtain AS-AS interactions. Second, by censoring all the other ASes, we obtain a smaller and intelligible Google matrix.

We saw that PageRank is a measure of centrality in terms of incoming links. In \graph, the PageRank centrality naturally captures the capacity of nodes to concentrate traffic. In order to study the nodes capacity to send traffic and its diffusive patterns, we make use of the reverse PageRank metric by calculating the stochastic complement of the inverted \graph. We note $G^*$ and $G_R^*$ the Google matrices associated to such a graph. 
In this case, the $(i,j)$ element of $G_R^*$ represents the probability, through direct and indirect links, that traffic arriving to node $j$ originates from $i$.
For the analysis, we don't show the diagonal values that represents self-loop traffic of an AS which is not of interest in this paper. To do so, we have to set diagonal value to 0 and re-normalize column-wise to get a stochastic matrix. 

\begin{figure}
    \centering
    \includegraphics[width = 4.5in]{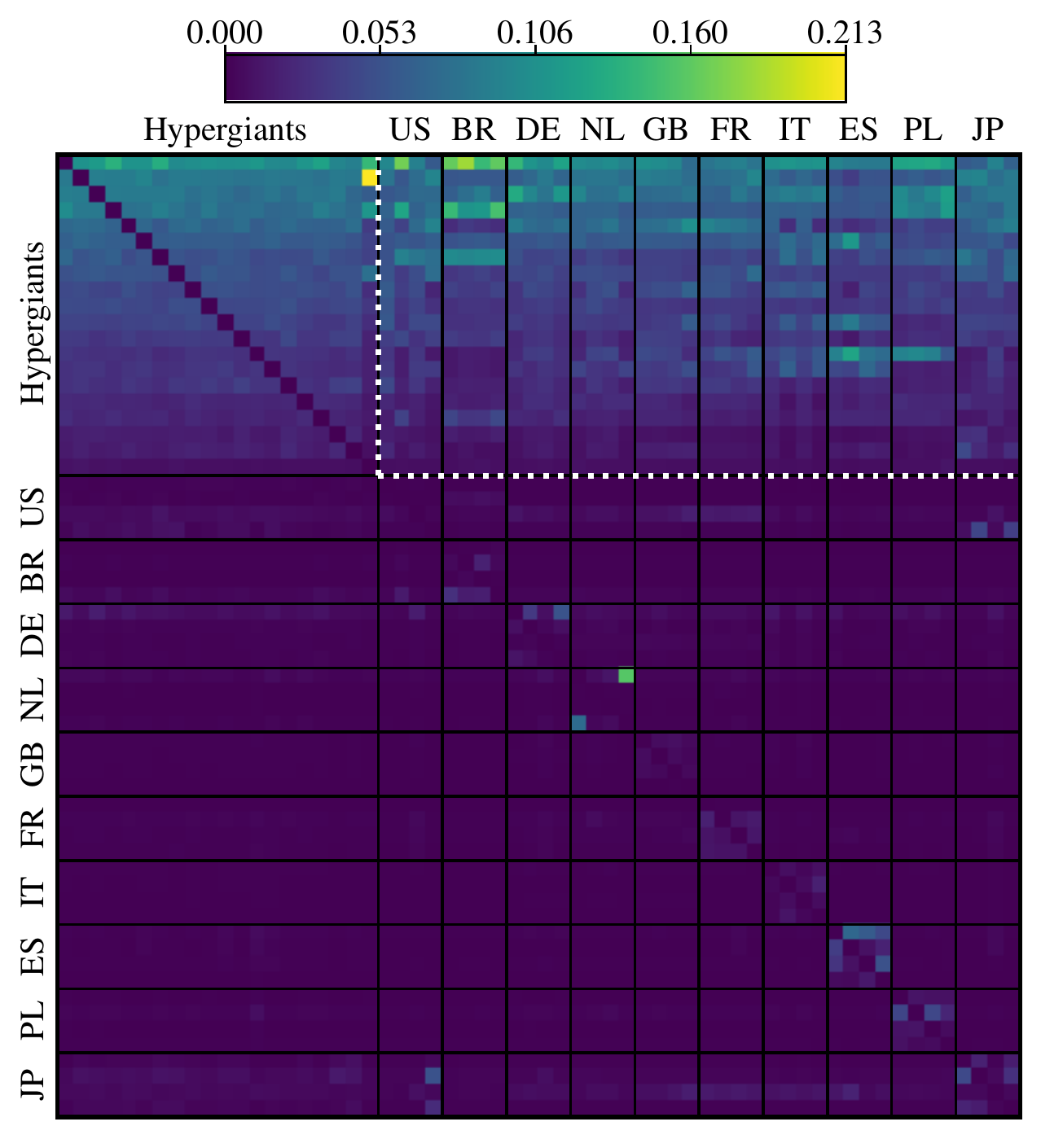}
    \caption{Reduced Google matrix $G_R^*$ for the reduced network of hypergiants and traffic receivers. {\it Hypergiants (resp. traffic receivers) are listed in the order of \autoref{tab:hypergiants_tab} (resp. \autoref{tab:catchers_tab}). The block encased in white dotted lines represents the hypergiant to traffic receivers diffusion part.}}
    \label{fig:Regomax_countries}
\end{figure}

The matrix of direct and indirect interactions $G_R^*$, represented in \autoref{fig:Regomax_countries}, shows two interesting properties. The first one is the block of strong links in the top right corner, identified with white dotted lines, that depicts the hypergiants diffusion to main regional ISPs. The concentration of links in this area confirms the grasp of hypergiants on regional ISPs. The second property is the block structures appearing on the matrix diagonal, highlighting local traffic exchanges enabled by IXPs. In particular, we report a strong link in the Netherlands between KPN-Netco and Joint-Transit, resulting from the fact that both players have their large port sizes solely at the same IXPs NL-ix and NL-ix2. This link captures the possible capacity that is offered by the physical network peering infrastructure to interconnect both ASes. However, there is currently is no AS path that reports both ASes in the BGP control plane. 
We can conclude this link shows that they benefit from interactions with the same third-party ASes in this case.



\section{Use case on COVID-19 demand for content}
\label{sec:covid}
    

In the previous sections, we saw that $i)$ hypergiants are at the core of the \graph\ and have a global reach and that $ii)$ the \graph\ structure reveals geographic proximity between hypergiants and regional eyeball ISPs. 

Hypergiant content delivery networks (CDN) connect to regional ISPs at IXPs to improve end-user experience \cite{chatzis2013benefits} and widen their reach. Covid-19 outbreak lead to a larger capacity demand from end-users \cite{feldmann2020lockdown}\cite{bottger2020internet}, forcing hypergiants to adjust their peering strategy by increasing total port size between \period\, as seen in the PeeringDB global port capacity evolution of \autoref{fig:portcapa_tot}. However it is not clear yet to what extent their proximity with regional ISPs has changed.
In this last part, we aim at identifying how hypergiants have increased or decreased their reach towards eyeballs using a reduced Google matrix analysis.

Therefore, we derive first the Google matrix of the full \graph. Next, we compute the Google matrix $G_R^*$ of the reduced network of ASes, this matrix encoding hypergiants diffusion to eyeballs. We compute $G_R^*\rvert^{d2}_{d1}$, the relative change of the elements of $G_R^*$ between two dates $d1$ and $d2$ given by
\begin{equation}
     G_R^*\rvert^{d2}_{d1}(i,j) = \frac{G_R^*\rvert_{d2}(i,j) - G_R^*\rvert_{d1}(i,j)}{G_R^*\rvert_{d1}(i,j)}.
\end{equation}

\begin{figure*}[h]
    \centering
    \includegraphics[scale=0.5, angle=-90]{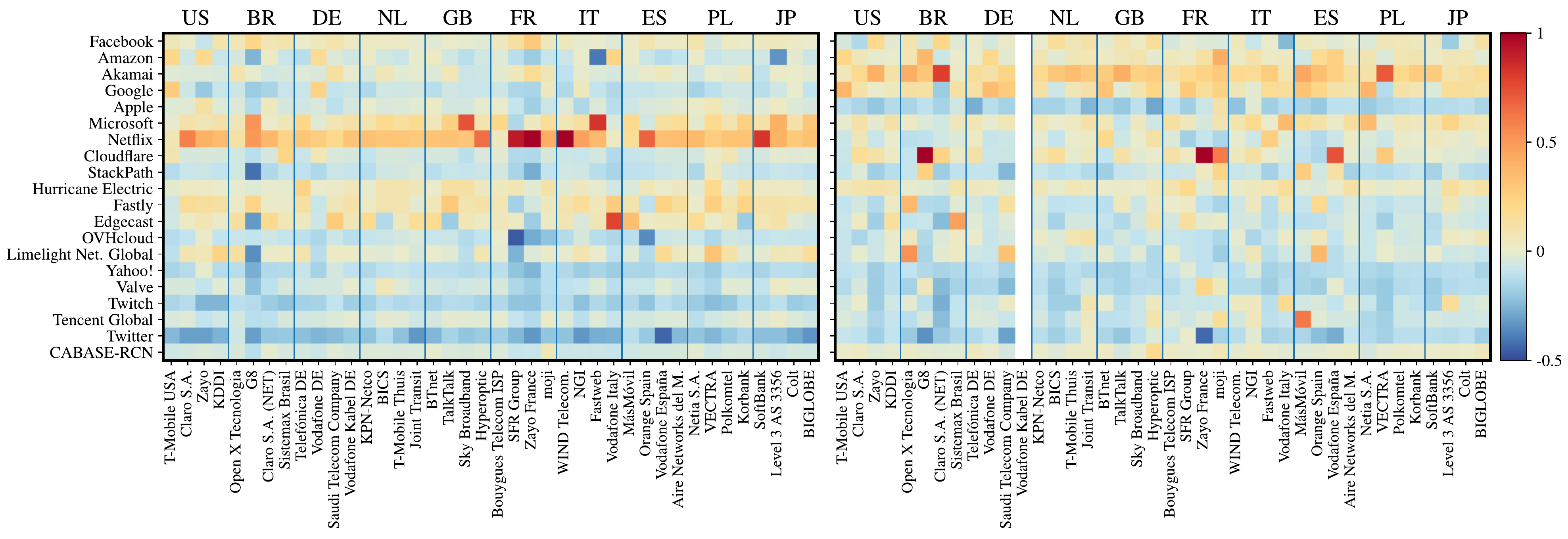}
    \caption{Relative time change of reduced Google matrices $G_R^*\rvert^{2020-06-01}_{2020-01-01}$ (left panel) and $G_R^*\rvert^{2021-01-01}_{2020-06-01}$ (right panel). {\it Only the top right part of $G_R^*$ is shown, corresponding to the hypergiants diffusion to traffic receivers part.}}
    \label{fig:Regomax_differences}
\end{figure*}

Results are shown in Figure~\ref{fig:Regomax_differences}.
For better visibility we limit the colormap to the interval ranging from -0.5 to 1.0, where -0.5 represents a 50\% decrease of a link from its initial value and 1.0 a 100\% increase. For the first period, the capped links are Netflix $\rightarrow$ Zayo France, Netflix $\rightarrow$ WIND Telecom with respective values 1.07, 1.03. For the second period, the capped links are Cloudflare $\rightarrow$ Zayo France, Cloudflare $\rightarrow$ G8 with values 1.34 and 1.09.

\FloatBarrier

During Covid outbreak, the biggest links increase are found for Netflix $\rightarrow$ (WIND Telecom, Zayo France, SFR Group, SoftBank), Microsoft $\rightarrow$ (Fastweb, Sky Broadband) and Edgecast $\rightarrow$ Vodafone Italy. The negative changes are OVHcloud $\rightarrow$ (SFR group, Orange Spain), Twitter $\rightarrow$ Vodafone Espa\~{n}a, StackPath $\rightarrow$ G8. We compute the sum of each line to determine which ASes have invested the most. 
Netflix is ahead by far, investing mainly in France, Italy, Japan and Great Britain, followed by Microsoft that invested in Great Britain, Italy and Japan, and then by Fastly in Poland, Italy, Great Britain, and finally Edgecast in Italy and Spain.
Netflix indeed reports an increase in capacity at IXPs to face a growing end-user demand during the outbreak \cite{NetflixOCA}. ASes that invested the least are Twitter, Twitch, Yahoo! and OVHcloud. 

Post-outbreak, the biggest links increase are found for Cloudflare $\rightarrow$ (Zayo France, G8, Vodafone Espa\~{n}a) and Akamai $\rightarrow$ (Claro SA, Vectra). The negative changes are notably Twitter $\rightarrow$ (Zayo France, G8, Saudi Telecom Company) and Apple $\rightarrow$ Hyperoptic.
The top investors are Akamai in Brazil, Poland, Great Britain, Spain and the Netherlands, followed by Google in Germany, Spain and France, and then Cloudflare in France and Brazil. ASes that invested the least are Twitter, Apple, Yahoo! and Valve.

\section{Related works}
\paragraph{IXP for Internet topology}
Internet topology being extensively measured and modeled over the last years, we restrict ourselves to the role IXPs played in the Internet topology. For an overview of AS-level studies, we refer the reader to \cite{roughan201110} and to the related works section of  \cite{nomikos2017re}.

Previous work of Ager et al. \cite{ager2012anatomy} shows that a single large European IXP presents more peering links than inferred for the Internet-wide AS-level topologies based on traceroute and BGP data. Access to the IXP traffic matrix and participants metadata allows the authors to characterize the diversity of the Internet ecosystem, revoking the classical AS tier classification. The global role of the same IXP is identified in \cite{chatzis2013benefits}, where authors argue that IXPs provide a good visibility of the Internet. \cite{camilo2012history} study the temporal evolution of an European IXP that gives insights on peering matrices, traffic growth, traffic imbalance and port utilization.

However, to the best of our knowledge, only a few studies have focused on the Internet-wide topology derived from public IXP datasets. An unweighted and undirected bipartite graph based on IXPs and their AS membership using data from PeeringDB \cite{PDB} and Packet Clearing House \cite{PCH} is proposed in \cite{nomikos2017re}. This allows a first study of ASes connectivity but do not contain information on traffic flows.

\paragraph{PeeringDB}
A first study of PeeringDB is presented in \cite{lodhi2014}. They show that PeeringDB entries are generally up-to-date and correct, and contain precious metadata on IXPs and their participants. They point out several biases, notably regarding AS business type diversities and their geographic distribution, but argue that PeeringDB gives a reasonable view of the peering ecosystem. A more recent overview of the dataset is presented in \cite{bottger2018looking}, where authors agglomerate AS port size at IXPs to introduce the \pc metric we leverage as well in this paper. Doing so, they successfully identify the hypergiants of the peering ecosystem. By combining this port size information to the bipartite model of \cite{nomikos2017re}, and introducing the direction of edges based on traffic imbalance metadata, we provide a weighted and directed graph that gives a coarse view of the traffic flows between the main protagonists of the peering ecosystem.

\paragraph{Google matrix}
PageRank \cite{BrinPage98} derived from the Google Matrix efficiently identifies popular pages of the hypertext-directed Web Graph. It has been applied to many other types of graphs \cite{Gleich2015}. In particular, combining PageRank and reverse PageRank study to the international trade network \cite{ermann2015google} allows to add new insights on money flows to the classical economic balance metric. Using this approach on our network, we are able to identify content hypergiants and main ISPs at the country level.

Stochastic complementation was proposed in \cite{meyer1989stochastic} to build a stochastic matrix for a small subset of nodes in a large network. This reduced stochastic matrix encodes information on the subset nodes interactions between themselves and over the whole network, leading to the differentiation of direct and indirect links. In our work we use this decomposition to obtain interactions between ASes, even though they do not share direct links. By doing so, we quantify the reach of hypergiants on the main regional ISPs and show the reaction of global content providers to the pandemic.

\section{Conclusion}
This paper proposes a novel Internet network model from PeeringDB database records, \graph, that offers a coarse but realistic picture of the overall capacity provisioned by ASes in the peering ecosystem. Its originality lies in the weighted and oriented edges which capture the port sizes and ASes \ir labels, respectively. 
From this model, we are able to identify key Internet players such as the state-of-the-art hypergiants and important regional eyeball networks present in the PeeringDB database. We show that it is possible, with a stochastic matrix representation of this graph, and its stochastic complementation for a reduced set of ASes, to extract quickly their capacity of interconnection offered the global physical public peering infrastructure. As a use-case, we propose a study that quantifies and identifies the links affected by the 2020 Covid-19 outbreak as captured by the PeeringDB ASes and IXPs evolution over time. 
Future works will investigate multi-layer networks \cite{bazzi2016community} or time-series similarity measures \cite{mori2015similarity} to better model the kinetic of the \graph.

\newpage

\appendix
\begin{table*}[t]
    \caption{IXP statistics for the 12 largest Louvain clusters of \graph\ of 2020-01-01.}
    \centering
\resizebox{0.98\textwidth}{!}{%
\begin{tabular}{@{}cccccc@{}}
\toprule
\begin{tabular}{c} Louvain\\Cluster\end{tabular}   & IXPs country distribution  & \begin{tabular}{c}\#Different \\IXP countries\end{tabular}   & Grouping interpretation   & Port capacity (\%) & \# of IXP (\%) \\ \midrule
0  & US: 74 | CA: 12 | CO: 2 | GU: 2 | SG: 2 & 13 & North America          & 18.5 & 14.6 \\
1  & DE: 22 | AT: 4 | CH: 4 | US: 2 | AE: 1  & 16 & Germanic countries     & 11.7 & 6.4  \\
2  & FR: 15 | GB: 11 | IE: 4 | US: 3 | AU: 2 & 16 & Western Europe         & 10.1 & 6.9  \\
4  & NL: 9 | BE: 1 | BJ: 1 | CA: 1 | IS: 1   & 6  & Netherlands            & 10.1 & 2.0  \\
7  & BR: 34 | RO: 4 | AO: 2 | KE: 2 | US: 2  & 11 & Brazil                 & 8.6  & 7.3  \\
8  & AR: 22 | JP: 13 | CL: 6 | HK: 2 | BH: 1 & 10 & South America, East Asia                      & 8.3  & 7.1  \\
5  & ID: 16 | TH: 7 | PH: 4 | HK: 3 | SG: 3  & 12 & South-East Asia           & 7.1  & 6.3  \\
10 & RU: 30 | UA: 12 | BG: 6 | KZ: 3 | KG: 2 & 12                        & Eastern Europe & 5.9                & 9.0                \\
3  & SE: 10 | NO: 7 | ES: 6 | FI: 4 | DK: 3  & 15 & Scandinavian countries & 5.0  & 6.7  \\
9  & AU: 21 | NZ: 6 | US: 4 | MY: 1          & 4  & Oceania Pacific        & 3.4  & 4.7  \\
6  & TZ: 5 | ZA: 5 | US: 4 | CA: 2 | CD: 2   & 36 & Africa                 & 2.8  & 7.3  \\
14 & PL: 11 | RO: 2 | PH: 1                  & 3  & Eastern Europe         & 2.3  & 2.0  \\ \bottomrule
\end{tabular}}
\label{tab:louvain_tab}
\end{table*}

\section{Reproducibility of our research}
We assess in this section the possibility of recreating our research. Based on ACM definitions \footnote{(\url{https://www.acm.org/publications/policies/artifact-review-and-badging-current})}, a work can either be repeatable (authors can reliably repeat their own computation), reproducible (an independent group can obtain the same results using the author's own artifacts) and replicable (an independent group can obtain the same results using artifacts which they develop completely independently).



\subsection{Repeatability}
Our results are repeatable. The dumps of PeeringDB are not subject to change and the processing is deterministic. 

\subsection{Reproducibility}
We will make available the generated datasets of \graph~ in \cite{loye2022peeringdb}, and plan to publicly share the source codes used for the dataset generation and analysis. The \graph\ is solely constructed from the public datasets of CAIDA. Our source codes rely on commonly used Python libraries, and our C\texttt{++} implementation of stochastic complementation \cite{loye2020pygomax} is cross-platform and self-contained thanks to CMake software development tool. We hope that these efforts will allow other researchers to reproduce and extend our results.

\subsection{Replicability}
The data we provide is either directly obtained from PeeringDB or from processing. We explicitly mention in this work the data acquired from processing and did our best to provide enough information to replicate this processing. We expect other researchers to be able to replicate our implementation or data analysis since it was performed with well-documented tools.

\section{Modularity clustering}\label{app:louvain}

 The Louvain method \cite{Blondel_2008} is a simple, unsupervised and easy-to-implement method for identifying communities in large networks. Moreover, it is known to be one of the most accurate community detection algorithms \cite{yang2016comparative}. This algorithm  relies on a greedy optimization method that minimizes the modularity metric of partitions of the network. The modularity measure for a bipartite network was introduced in \cite{barber2007modularity}, and a Louvain implementation using this modularity is proposed in \cite{feng2020improving}. 
Since the bipartite modularity is defined for undirected networks, we apply Louvain to the undirected \graph~of weighted adjacency matrix $A = W + W^T$. A link of this network corresponds to the port size an AS possesses at an IXP. 


Louvain method is known to fail in detecting small communities, but we are interested in identifying larger ones. 
 Studying the distribution of IXPs country in Louvain clusters, we observe that the algorithm grouped countries with strong relations or geographical proximity, as underlined in Table~\ref{tab:louvain_tab}.
Clusters grouping Northern American, Brazilian, Asian-Pacific, Scandinavian or Eastern European IXPS are observed consistently throughout our testings.

\newpage
\bibliographystyle{plain}
\bibliography{references}

\end{document}

%% file: info_ratio_tab.tex
\begin{tabular}{@{}lcc@{}}
\toprule
\ir & \pc   & \multicolumn{1}{c}{Count} \\ 
Categories                         & Proportion (\%) & Proportion (\%)           \\ \midrule
Balanced    (B)                       & 26.12           & 31.69                     \\
Heavy Inbound  (HI)                    & 4.72           & 6.80                      \\
Heavy Outbound (HO)                    & 19.68           & 3.68                      \\
Mostly Inbound (MI)                    & 22.45           & 30.69                     \\
Mostly Outbound (MO)                    & 21.34           & 10.86                     \\
Not Disclosed                      & 5.69            & 16.28                     \\ \bottomrule
\end{tabular}

%% file: stability_tab.tex
\begin{tabular}{@{}llllll@{}}
\toprule
                                     & IR    & PR & $\Delta$ PR & rPR & $\Delta$ rPR \\ \midrule
Facebook                          & HO  & 124       & 1015                & 4                & 0                          \\
Akamai                            & HO  & 172      & 1330               & 7                & 1                          \\
StackPath (Highwinds)             & HO  & 314      & 2214               & 16                & 2                          \\
Netflix                           & HO  & 255      & 1853               & 11               & 0                          \\
Apple                            & MO & 34       & 29                 & 9               & 0                          \\
Google                           & MO & 27       & 17                 & 8                & 1                          \\
Microsoft                        & MO & 41       & 32                 & 10                & 0                          \\
Cloudflare                       & MO & 56       & 33                 & 12               & 2                          \\
Amazon.com                       & B        & 4        & 0                  & 5               & 1                          \\
Hurricane Electric               & B        & 7       & 2                  & 19               & 2                          \\
Core-Backbone                    & B        & 70      & 4                  & 96              & 2                         \\
Telefónica Germany               & B        & 72       & 8                  & 100              & 4                          \\
KPN-Netco                        & MI  & 32       & 5                  & 150              & 58                         \\
SoftBank Corp.                   & MI  & 69       & 3                 & 272              & 131                        \\
Vodafone Germany                 & MI  & 86       & 14                 & 360              & 176                        \\
Saudi Telecom Company            & MI  & 104       & 12                 & 390              & 176                        \\
Telekomunikasi Indonesia         & HI   & 153       & 18                 & 2041             & 5412                       \\
Charter Communications           & HI   & 175      & 11                 & 2233             & 5431                       \\
Open X Tecnologia	             & HI   & 103      & 15                 & 1539             & 5149                       \\
OPTAGE                           & HI   & 334      & 15                 & 3230             & 5669                       \\ \bottomrule
\end{tabular}

%% file: PRCR_comp_tab.tex
\resizebox{0.7\textwidth}{!}{%
\begin{tabular}{@{}llll@{}}
\toprule
   & \begin{tabular}[c]{@{}l@{}}PageRank\\ \graph \end{tabular} & \begin{tabular}[c]{@{}l@{}}PageRank \\ graph of \cite{nomikos2017re}\end{tabular} & \begin{tabular}[c]{@{}l@{}}Reverse PageRank\\ \graph \end{tabular} \\ \midrule
1  & IX.br São Paulo   & IX.br São Paulo    & IX.br São Paulo   \\
2  & DE-CIX Frankfurt               & AMS-IX                       & DE-CIX Frankfurt   \\
3  & AMS-IX                         & DE-CIX Frankfurt                & AMS-IX             \\
4  & \textbf{Amazon}                      & LINX LON1                       & \textbf{Facebook}       \\
5  & LINX LON1                    & EPIX.Katowice                  & \textbf{Amazon}        \\
6  & NAPAfrica IX J.       & Mumbai IX                      & LINX LON1           \\
7  & \textbf{Hurricane Electric}         & NAPAfrica IX J.   & \textbf{Akamai}       \\
8  & Equinix Singapore               & France-IX Paris                & \textbf{Google}        \\
9  & NL-ix                         & SIX Seattle                     & \textbf{Apple}       \\
10 & SIX Seattle                    & NL-ix                           & \textbf{Microsoft}          \\
11 & Equinix Ashburn               & IX.br Rio de Janeiro   & \textbf{Netflix}            \\
12 & EPIX.Katowice                 & TorIX                          & \textbf{Cloudflare}        \\
13 & IX.br Rio de Janeiro   & EPIX.Warszawa-KIX               & NL-ix              \\
14 & Equinix Chicago                 & LINX LON2                      & Equinix Singapore  \\
15 & Netnod Stockholm        & Equinix Ashburn             & NAPAfrica IX J. \\ \bottomrule
\end{tabular}}

%% file: geographical_classif_tab.tex
\begin{tabular}{@{}ccccc@{}}
\toprule
Country & Precision & Recall & F1-Score & Support \\ \midrule
US & 0.88      & 0.89   & 0.88     & 1360    \\
CA & 0.93      & 0.80   & 0.86     & 297     \\
BR & 0.99      & 0.99   & 0.99     & 1302    \\
DE & 0.75      & 0.86   & 0.80     & 484     \\
NL & 0.73      & 0.81   & 0.77     & 288     \\
GB & 0.90      & 0.76   & 0.82     & 440     \\
FR & 0.92      & 0.86   & 0.89     & 308     \\
IT & 0.94      & 0.91   & 0.93     & 228     \\
ES & 0.96      & 0.90   & 0.93     & 114     \\
RU & 0.90      & 0.91   & 0.90     & 290     \\
PL & 0.98      & 0.96   & 0.97     & 421     \\
CN & 0.79      & 0.48   & 0.60     & 56      \\
IN & 0.99      & 0.98   & 0.99     & 438     \\
JP & 0.92      & 0.96   & 0.94     & 227 \\ \bottomrule
\end{tabular}


%% file: hypergiants_tab.tex
\begin{tabular}{@{}clcl@{}}

\toprule
Rank & Name & Rank & Name             \\ \midrule
1    & \textbf{Facebook} & 11 & Fastly        \\
2    & \textbf{Amazon}   & 12 & \textbf{Edgecast}          \\
3    & \textbf{Akamai}   & 13 & \textbf{OVHcloud} \\
4    & \textbf{Google}   & 14 & \textbf{Limelight Networks Global}  \\
5    & \textbf{Apple}    & 15 & \textbf{Yahoo!}         \\
6    & \textbf{Microsoft} & 16 & Valve Corporation      \\
7    & \textbf{Netflix}   & 17 & \textbf{Twitch}       \\
8 & \textbf{Cloudflare} & 18 & Tencent Global \\
9 & StackPath & 19 & \textbf{Twitter} \\
10 & \textbf{Hurricane Electric} & 20 & CABASE-RCB \\ \bottomrule

\end{tabular}

%% file: coverage_horizontal.tex


\resizebox{0.98\textwidth}{!}{%
\begin{tabular}{@{}lcccccccccccccc@{}}
\toprule
Selection procedure   & US    & CA    & BR    & DE    & NL    & GB    & FR    & IT    & ES    & RU    & PL    & CN    & IN    & JP    \\ \midrule
APNIC       & 96.61 & 99.49 & 94.57 & 99.36 & 99.67 & 99.39 & 99.67 & 99.55 & 99.72 & 98.24 & 99.11 & 99.85 & 99.24 & 99.69 \\
PeeringDB and APNIC       & 32.46 & 86.88 & 53.44 & 92.84 & 86.55 & 87.44 & 57.27 & 60.28 & 56.75 & 40.25 & 37.25 & 56.44 & 9.96 & 92.12 \\
PeeringDB and APNIC top4 & 18.78 & 52.98 & 29.75 & 60.38 & 66.01 & 61.85 & 51.75 & 53.95 & 49.52 & 30.09 & 26.64 & 56.07 & 4.34 & 63.54 \\
Top4 PR ISP/ND      & \textbf{13.67} & 3.44  & 13.84 & 24.32 & 7.81  & \textbf{43.98} & \textbf{28.09} & 13.23 & \textbf{45.07} & 0.71  & \textbf{15.82} & 0.01  & \textbf{3.32}  & 14.54 \\
Traffic receivers      & \textbf{9.57}  & 3.44  & 13.87 & 29.38 & 32.77 & \textbf{42.02} & \textbf{28.04} & \textbf{50.12} & \textbf{46.38} & 0.01  & \textbf{19.65} & 0.00  & \textbf{3.33}  & 21.96 \\ \bottomrule
\end{tabular}}

%% file: catchers_apnic_tab.tex
\begin{tabular}{lccccS}
\toprule
\multirow{2}{*}{Name} &
  \multicolumn{2}{c}{PeeringDB}  &
  \multicolumn{3}{c}{APNIC} \\
  \cpartline{2-3} \cpartline{4-6}
  & {CC} & {Rank} & {CC} & {Rank} & {EUMS} \\
  \midrule
T-Mobile USA          & US        & 1         & US             & 2           & 9.53          \\
Claro S.A.            & US        & 2         & BR             & 24          & 0.16          \\
Zayo                  & US        & 3         & US             & 44          & 0.04          \\
KDDI                  & US        & 4         & JP             & 1           & 19.76         \\
Open X Tecnologia     & BR        & 1         & BR             & 40          & 0.0           \\
G8                    & BR        & 2         & BR             & 35          & 0.05          \\
Claro S.A. (NET)      & BR        & 3         & BR             & 1           & 13.82         \\
Sistemax Brasil       & BR        & 4         & BR             & 40          & 0.0           \\
Telefónica DE         & DE        & 1         & DE             & 3           & 10.14         \\
Vodafone DE           & DE        & 2         & DE             & 2           & 11.61         \\
Saudi Telecom Company & DE        & 3         & SA             & 17          & 0.0           \\
Vodafone Kabel DE     & DE        & 4         & DE             & 5           & 7.63          \\
KPN-Netco             & NL        & 1         & NL             & 1           & 29.85         \\
BICS                  & NL        & 2         & CN             & 39          & 0.0           \\
T-Mobile Thuis        & NL        & 3         & NL             & 6           & 2.92          \\
Joint Transit         & NL        & 4         &                &             &               \\
BTnet                 & GB        & 1         & GB             & 3           & 16.77         \\
TalkTalk              & GB        & 2         & GB             & 5           & 5.77          \\
Sky Broadband         & GB        & 3         & GB             & 1           & 19.21         \\
Hyperoptic            & GB        & 4         & GB             & 20          & 0.27          \\
Bouygues Telecom ISP  & FR        & 1         & FR             & 4           & 13.65         \\
SFR Group             & FR        & 2         & FR             & 3           & 14.37         \\
Zayo France           & FR        & 3         & FR             & 21          & 0.02          \\
moji                  & FR        & 4         & FR             & 23          & 0.0           \\
WIND Telecom.         & IT        & 1         & IT             & 3           & 17.02         \\
NGI                   & IT        & 2         & IT             & 8           & 1.37          \\
Fastweb               & IT        & 3         & IT             & 4           & 10.44         \\
Vodafone Italy        & IT        & 4         & IT             & 2           & 21.29         \\
MásMóvil              & ES        & 1         & ES             & 4           & 6.35          \\
Orange Spain          & ES        & 2         & ES             & 2           & 24.92         \\
Vodafone España       & ES        & 3         & ES             & 3           & 13.63         \\
Aire Networks del M.  & ES        & 4         & ES             & 10          & 1.48          \\
Netia S.A.            & PL        & 1         & PL             & 5           & 3.92          \\
VECTRA                & PL        & 2         & PL             & 6           & 3.38          \\
Polkomtel             & PL        & 3         & PL             & 2           & 12.27         \\
Korbank               & PL        & 4         & PL             & 26          & 0.08          \\
SoftBank              & JP        & 1         & JP             & 2           & 19.23         \\
Level 3 AS 3356       & JP        & 2         & US             & 39          & 0.09          \\
Colt                  & JP        & 3         & DE             & 30          & 0.09          \\
BIGLOBE               & JP        & 4         & JP             & 8           & 2.73          \\ \bottomrule
\end{tabular}